\documentclass[traditabstract]{aa}
\usepackage{txfonts}
\usepackage{graphicx}
\usepackage{subfig}
\usepackage{natbib}
\bibpunct{(}{)}{;}{a}{}{,}

\usepackage[colorlinks=true, citecolor=blue, linkcolor=blue, breaklinks=true]{hyperref}
\newcommand{\hi } {{\rm H}\,{\small\rm I} \,}

\begin{document}

\title{Structure and Dynamics of Giant Low Surface Brightness Galaxies}
\author{Federico Lelli\inst{1,}\inst{3}
\and Filippo Fraternali\inst{1}
\and Renzo Sancisi\inst{2,}\inst{3}}

\institute{Department of Astronomy, University of Bologna, via Ranzani 1, 40127, Bologna, Italy\\
\email{lelli@astro.rug.nl}
\and INAF - Astronomical Observatory of Bologna, via Ranzani 1, 40127, Bologna, Italy
\and Kapteyn Astronomical Institute, Postbus 800, 9700 AV, Groningen, The Netherlands}

\date{Received 7 December 2009 / Accepted 25 February 2010}

\abstract{
Giant low surface brightness (GLSB) galaxies are commonly thought to be massive,
dark matter dominated systems. However, this conclusion is based on highly uncertain rotation curves.
We present here a new study of two prototypical GLSB galaxies: Malin~1 and NGC~7589.
We re-analysed existing \hi observations and derived new rotation curves, which
were used to investigate the distributions of luminous and dark matter in these galaxies.

In contrast to previous findings, the rotation curves of both galaxies show a steep
rise in the central parts, typical of high surface brightness (HSB) systems.
Mass decompositions with a dark matter halo show that baryons may dominate the dynamics
of the inner regions. Indeed, a ``maximum disk'' fit gives stellar mass-to-light ratios
in the range of values typically found for HSB galaxies.

These results, together with other recent studies,
suggest that GLSB galaxies are systems with a double structure:
an inner HSB early-type spiral galaxy and an outer extended LSB disk.

We also tested the predictions of MOND: the rotation curve of NGC~7589 is reproduced well,
whereas Malin~1 represents a challenging test for the theory.
}

\keywords{dark matter -- Galaxies: structure -- Galaxies: kinematics and dynamics -- Galaxies: individual: Malin 1 -- Galaxies: individual: NGC 7589 -- Methods: data analysis}
\titlerunning{Structure and Dynamics of Giant Low Surface Brightness Galaxies}
\authorrunning{Lelli et al.}

\maketitle

\section{Introduction}

Low surface brightness (LSB) galaxies span a wide range of galaxy sizes, masses and morphologies, from dwarf spheroidals and dwarf irregulars to medium-size late-type disks (Sc-Sd) and to bulge-dominated early-type spirals (Sa-Sb). According to \citet{Beijersbergen1999}, LSB spirals form a LSB Hubble sequence, parallel to the classical HSB one.

Giant low surface brightness (GLSB) galaxies are usually considered as extreme cases of early-type LSB spirals \citep{Beijersbergen1999}. They have exceedingly extended LSB disks, with scale lengths ranging from $\sim$10 to $\sim$50 kpc, and central luminous components resembling the bulges of HSB spirals (e.g. \citealt{Bothun1987, Sprayberry1995}). They are massive systems, with $L \sim L_{*}$. Like the ordinary LSB spirals, GLSB galaxies have low \hi surface densities \citep{Pickering1997}, but they are among the most gas rich galaxies known, with $M_{\rm{\hi }}\approx 10^{10} M_{\odot}$ \citep{Matthews2001b}.

Few dynamical studies exist on GLSB galaxies. \citet{Pickering1997} studied four GLSB galaxies and found slowly rising rotation curves, similar to those of late-type LSB disks (e.g. \citealt{deBlok1997,DeNaray2006}). They concluded that GLSB galaxies are ``the first examples of galaxies that are both massive and dark matter dominated''. However, they warned that their rotation curves are highly uncertain, due to the low signal-to-noise ratio and the low spatial resolution of the observations.

A slowly rising rotation curve in the presence of an inner concentration of light, as seen in those galaxies (see top panels of figure \ref{fig:deco1}), is in marked contrast with the rule that there is a close correlation between the concentration of light and the shape of the rotation curve (\citealt{Sancisi2004} and references therein). For this reason, \citet{Sancisi2007} started a re-analysis of the 21-cm data from \citet{Pickering1997}. This paper concludes that preliminary work with the study of the structure and dynamics of two GLSB galaxies: Malin~1 and NGC~7589.

\section{\label{due}\hi Data Analysis}

We re-analyzed the \hi datacubes of Malin~1 and NGC~7589, obtained with the VLA by \citet{Pickering1997}. The angular resolutions are low: $20''.9\times20''.8$ for Malin~1 (corresponding to $\sim 32$ kpc for an angular-diameter distance $D_{\rm{A}} = 322$ Mpc) and $21''.4 \times 17''.3$ for NGC~7589 ($\sim  12 \times 10$ kpc for $D_{\rm{A}}= 123$ Mpc). We used the Groningen imaging processing system (GIPSY) \citep{vanderHulst1992}. The main physical properties of both galaxies are listed in table \ref{tab:HIres}.

Total \hi maps were obtained by adding the channel maps with line emission (from 24581 km~s$^{-1}$ to 24955 km~s$^{-1}$ for Malin~1 and from 8566 km s$^{-1}$ to 9157 km~s$^{-1}$ for NGC~7589). In order to define the areas of emission in each channel map, we used masks, which were obtained by smoothing the original datacubes to lower spatial resolutions. We inspected by eye each channel map overlaying different masks. We obtained the best results by smoothing to $40''$ and $50''$ and clipping at $2 \sigma_{\rm{s}}$ and $3 \sigma_{\rm{s}}$ for Malin~1 and NGC~7589 respectively ($\sigma_{\rm{s}}$ is the rms noise in the smoothed cube). The resulting total \hi maps are shown in figure \ref{fig:Maps}.

\begin{table}[thbp]
\caption{Physical Properties of Malin~1 and NGC~7589.}
\label{tab:HIres}
\begin{center}
\begin{tabular}{l c c c}
\hline
\hline
Property & Malin 1 & NGC 7589 & Ref.\\
\hline
Morphological Type    & SB0/a     & SAB(rs)a        & 1, 2\\
AGN activity          & LINER   &  BLAGN          & 1, 2\\
M$_{\rm{B}}$ (mag)    & $-22.1 \pm 0.4 $ & $-20.4 \pm 0.2$ & 3\\
Redshift              & $0.0826 \pm 0.0017$ & $0.0298 \pm 0.0017$ & 4\\
$D_{\rm{L}}$ (Mpc)         & $377 \pm 8$ & $130 \pm 8$ & 4\\
$D_{\rm{A}}$ (Mpc)         & $322 \pm 6$ & $123 \pm 7$ & 4\\
\hi flux (Jy km s$^{-1}$)  & $2.0 \pm 0.2$ & $3.8 \pm 0.3$ & 4\\
$M_{\rm{\hi }}$ ($10^{10} M_{\odot}$)    & $6.7 \pm 1.0$ & $1.5 \pm 0.3$ & 4\\
SFR$_{\rm{FUV}}$ ($M_{\odot}$ yr$^{-1}$) & 1.2 & 0.7 & 5 \\
SFR$_{\rm{NUV}}$ ($M_{\odot}$ yr$^{-1}$) & 2.5 & 1.1 & 5 \\
\hline
\hline
\end{tabular}
\tablebib{
(1)~\citet{Barth2007}; (2)~Nasa Extra-galactic Database (NED); (3)~\citet{Pickering1997} give M$_{\rm{V}} = -22.9 \pm 0.4$ for Malin~1 and M$_{\rm{R}} = -21.9 \pm 0.2$ for NGC~7589; \citet{Impey1989} give $B - V = 0.8$ for Malin~1; \citet{Galaz2006} give $B-R = 1.5$ for NGC~7589; (4)~This work; (5)~\citet{Boissier2008}.
}
\end{center}
\end{table}

From the masked data, we calculated the \hi fluxes (correcting for primary beam attenuation).
For Malin 1, we obtained $S_{\rm{tot}} = 2.0 \pm 0.2$ Jy~km~s$^{-1}$, consistent within the errors with
the value derived by \citet{Pickering1997} ($S_{\rm{tot}} = 2.5 \pm 0.2$ Jy~km~s$^{-1}$) and the single-dish measurement by \citet{Matthews2001b} ($S_{\rm{tot}}=1.80 \pm 0.50$ Jy~km~s$^{-1}$).
For NGC 7589, we obtained $S_{\rm{tot}} = 3.8 \pm 0.3$ Jy~km~s$^{-1}$, whereas \citet{Pickering1997} derived $S_{\rm{tot}}=2.7 \pm 0.3$ Jy~km~s$^{-1}$. The differences are probably due to the different masks used; the values are consistent within $2\sigma$. Single-dish observations gave intermediate values, consistent with our value within about $1\sigma$: $S_{\rm{tot}}=2.96 \pm 0.51$ Jy~km~s$^{-1}$ \citep{Matthews2001b} and $S_{\rm{tot}}=3.07 \pm 0.36$ Jy~km~s$^{-1}$ \citep{Springob2005} (this latter value is without the correction for \hi self absorption).
\hi masses were estimated using the luminosity distance $D_{\rm{L}}$, given by:
\begin{equation}
\centering
D_{\rm{L}} = \frac{c}{H_{0}} \, [z + 0.5 \times (1 - q_{0}) \, z^{2}].
\end{equation}
We assumed $H_{0}=70$ km~sec$^{-1}$~Mpc$^{-1}$, $\Omega_{\rm{m}}=0.27$ and $\Omega_{\rm{\Lambda}}=0.73 \,$ ($q_{0}=-0.59$). Since we are dealing with low redshift galaxies, the exact value of $q_{0}$ affects our results only slightly. The redshift was calculated as $z=V_{\rm{sys}}/c$, where $V_{\rm{sys}}$ is the galaxy systemic velocity derived from the tilted ring analysis (see section \ref{RC}). Our results are listed in table \ref{tab:HIres}.

\subsection{\label{velo}\hi velocity fields}

The derivation of the velocity fields is not straightforward because of beam-smearing effects, which are particularly severe here because of the low spatial resolution. Indeed, especially in the inner regions, the 21-cm line profiles are unusually broad and asymmetric. A Gaussian fit or an intensity-weighted mean (IWM), as used by \citet{Pickering1997}, give velocities far from the profile peaks and biased towards the systemic velocity. This leads to velocity fields with systematically underestimated inner velocity gradients and consequently to slowly rising rotation curves. This is illustrated by figure~7 of \citet{Pickering1997}, which shows the rotation curves of Malin~1 and of NGC~7589 overlaid on slices of the datacubes along the major axes of the galaxies. Clearly, the inner points of the rotation curves are at lower velocities than the emission peaks, suggesting that they are underestimated. The importance of beam-smearing effects is demonstrated by the galaxy models presented in section \ref{sub:models}.

\begin{table}[thbp]
\caption{Kinematical Parameters of Malin~1 and NGC~7589}
\label{tab:kinparams}
\begin{center}
\begin{tabular}{l c c}
\hline
\hline
Parameter & Malin 1 & NGC 7589 \\
\hline
$ \alpha_{\rm{opt}} (\rm{J}2000.0)$ & $12^{h} 36^{m} 59^{s}.4 \pm 1^{s}.8$ & $23^{h} 18^{m} 15^{s}.7 \pm 1^{s}.8$ \\
$ \delta_{\rm{opt}} (\rm{J}2000.0)$ & $14^{\circ} 19' 49''.3 \pm 1''.8$      &  $00^{\circ} 15' 40''.2 \pm 1''.8 $\\
$ \alpha_{\rm{kin}} (\rm{J}2000.0)$ & $12^{h} 36^{m} 59^{s}.2 \pm 0^{s}.1$ & $23^{h} 18^{m} 15^{s}.5 \pm 0^{s}.2$ \\
$ \delta_{\rm{kin}} (\rm{J}2000.0)$ & $14^{\circ} 19' 51''.3 \pm 1''.8$      &  $00^{\circ} 15' 37''.3 \pm 2''.5 $\\
$V_{\rm{sys}}$ (km s$^{-1}$) & $24766.7 \pm 4.0$ & $8926.4 \pm 3.8$ \\
Position Angle            & see figure \ref{fig:pvMalin1} & $307^{\circ}.7 \pm 1^{\circ}$.1 \\
Inclination Angle         & $38^{\circ} \pm 3^{\circ}$  & $58^{\circ} \pm 3^{\circ}$ \\
\hline
\hline
\end{tabular}
\tablefoot{The optical centres ($\alpha_{\rm{opt}}, \delta_{\rm{opt}}$) are taken from NED. The inclination angles are taken from \citet{MooreParker2006} for Malin~1 and from \citet{Pickering1997} for NGC~7589.}
\end{center}
\end{table}

\begin{figure*}
\centering
\includegraphics[width=17cm]{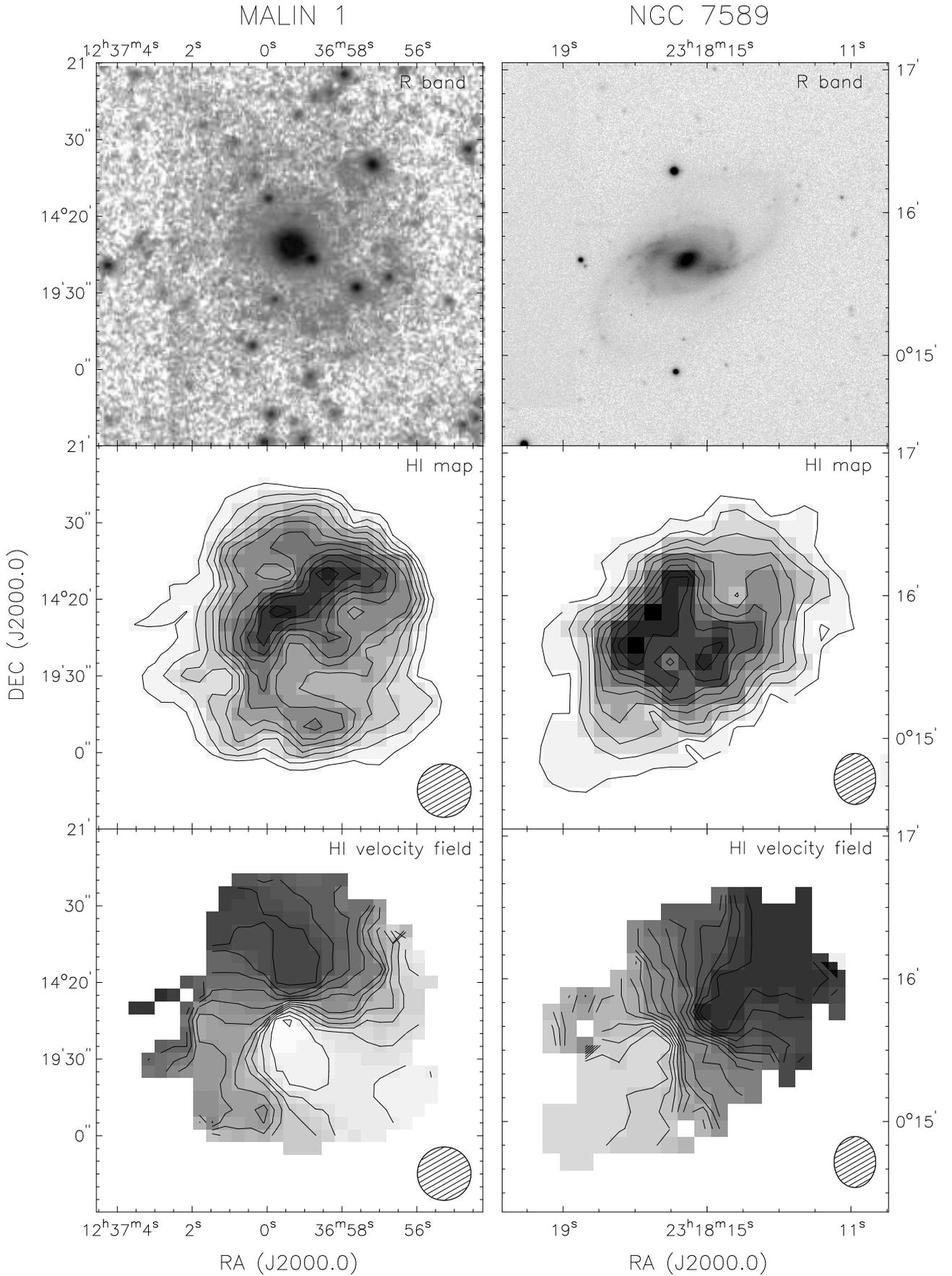}
\caption{\textit{Left Panels:} Malin~1. \textit{Top}: R-band image \citep{MooreParker2006}. \textit{Middle}: total \hi map. Contours range from $0.9 \times 10^{20}$ to $5.4 \times 10^{20}$ atoms~cm$^{-2}$ with steps of $0.45 \times 10^{20}$ atoms~cm$^{-2}$. \textit{Bottom}: \hi velocity field. The receding side is in dark-grey. Contours range from $24623$ km~s$^{-1}$ to $24887$ km~s$^{-1}$ with steps of $24$ km~s$^{-1}$ (2 times the channel spacing). \textit{Right Panels:} NGC~7589. \textit{Top}: R-band image \citep{Galaz2006}. \textit{Middle}: total \hi map. Contours range from $1.1 \times 10^{20}$ to $9.9 \times 10^{20}$ atoms~cm$^{-2}$ with steps of $1.1 \times 10^{20}$ atoms~cm$^{-2}$. \textit{Bottom}: \hi velocity field. The receding side is in dark-grey. Contours range from $8772$ km~s$^{-1}$ to $9080$ km~s$^{-1}$ with steps of $22$ km~s$^{-1}$ (the channel spacing).}
\label{fig:Maps}
\end{figure*}

We have derived new velocity fields with a different technique to minimize beam-smearing effects. We selected velocities near the peaks of the emission, estimating an IWM velocity from the upper part of the line profile (above $\sim 75\%$ of the peak intensity) and neglecting the broad wings.
This is like taking the profile peak velocities, but with the advantage of less noise in the resulting velocity field. In the outer regions of the galaxies, where the line profiles are more regular and symmetric, we used a Gaussian fit. The resulting velocity fields are shown in figure~\ref{fig:Maps}.

The left panels of figure \ref{fig:Maps} show the R-band image of Malin~1 from \citet{MooreParker2006} (top), the total \hi map (middle) and the \hi velocity field (bottom). The \hi distribution is approximately as extended as the optical LSB disk. The outer \hi radius measures $\sim 64''$, which corresponds to $\sim 100$ kpc. On the southern side, a spiral arm is visible. It extends for more than 80 kpc and apparently it is the gaseous counterpart of the spiral arm detected by \citet{MooreParker2006}. The velocity field shows a strong variation of the position angle of the major axis with radius, both in the approaching and in the receding sides of the galaxy. Such behaviour is usually attributed to the presence of a warp. Overall, the velocity field is regular and symmetric.

The right panels of figure \ref{fig:Maps} show the R-band image of NGC~7589 from \citet{Galaz2006} (top), the total \hi map (middle) and the \hi velocity field (bottom). The \hi distribution is as extended as the optical LSB disk. The outer \hi radius measures $\sim 80''$, which corresponds to $\sim 48$ kpc. The \hi distribution follows that of the light, with two main concentrations (North-East and South-West) and two faint outer spiral arms. The velocity field is regular and symmetric.

\begin{figure}
\centering
\includegraphics[width=7.3 cm]{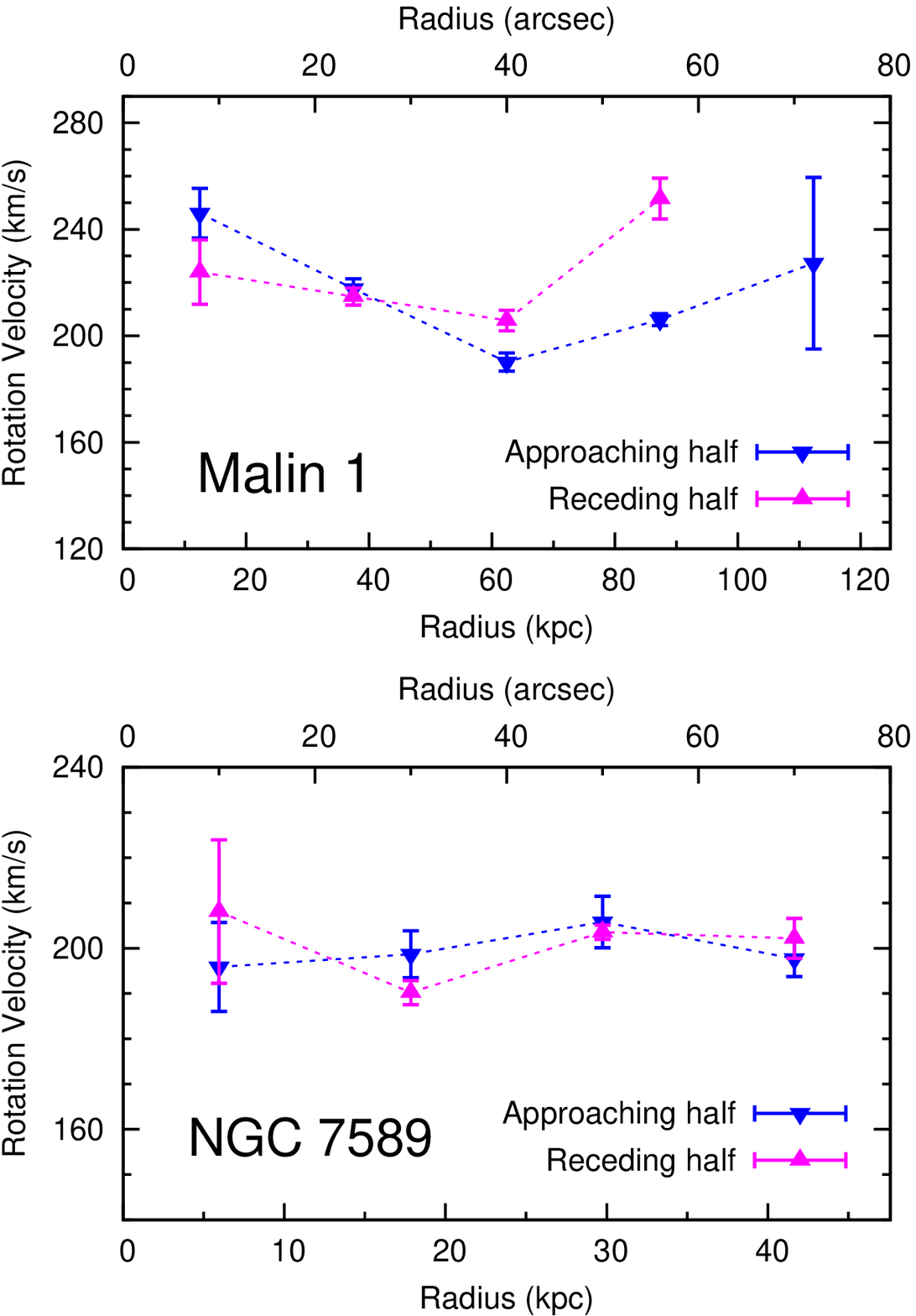}
\caption{\hi rotation curves for Malin 1 (\textit{top}) and NGC 7589 (\textit{bottom}), derived separately for the receding (up-triangles) and approaching (down-triangles) halves.}
\label{fig:vrotAppRec}
\end{figure}

\begin{figure*}[tbh]
\begin{minipage}{0.5\textwidth}
\centering
\includegraphics[width=10 cm]{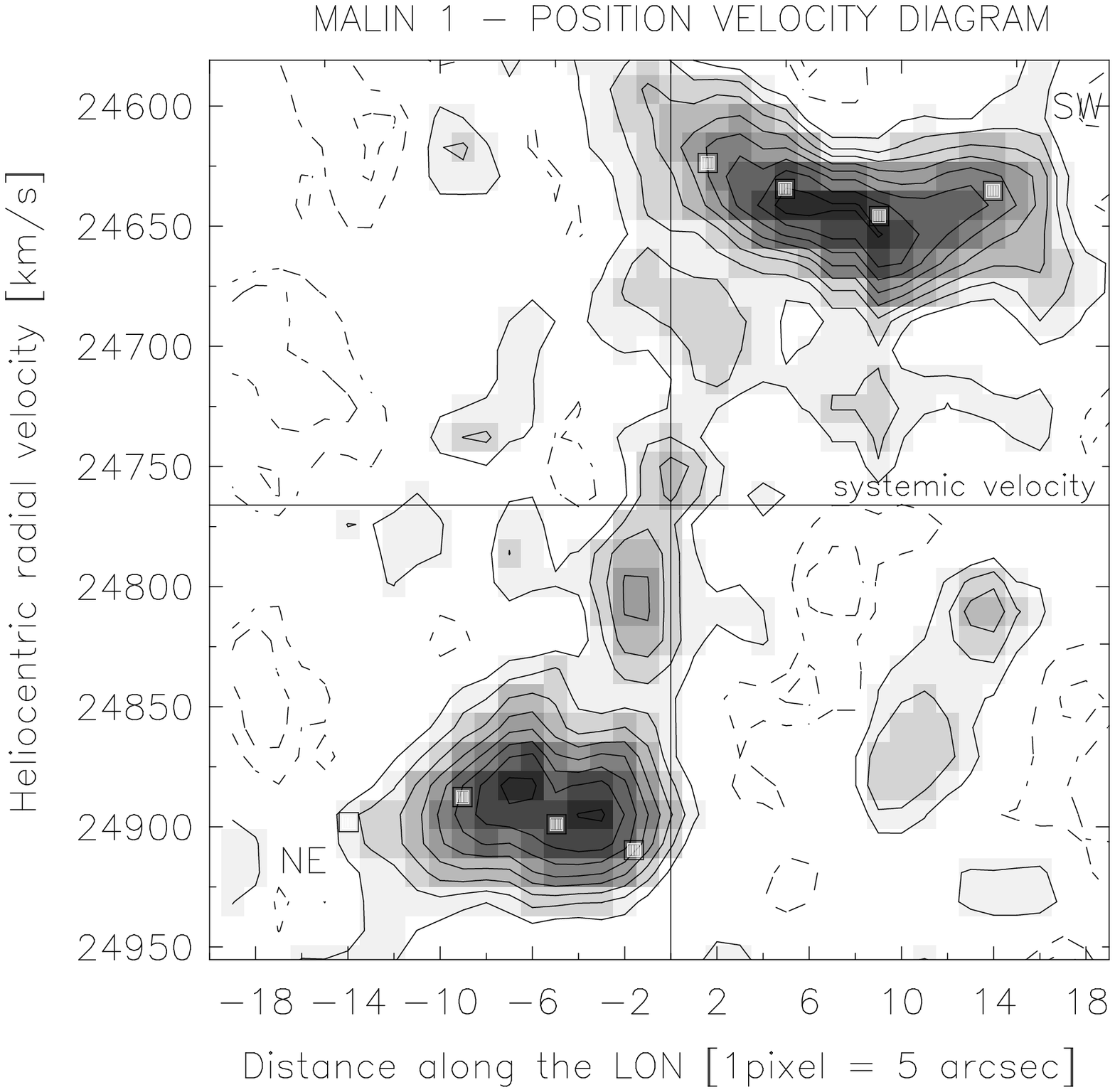}
\end{minipage}
\begin{minipage}{0.5\textwidth}
\centering
$\begin{array}{c}
\includegraphics[width=7 cm]{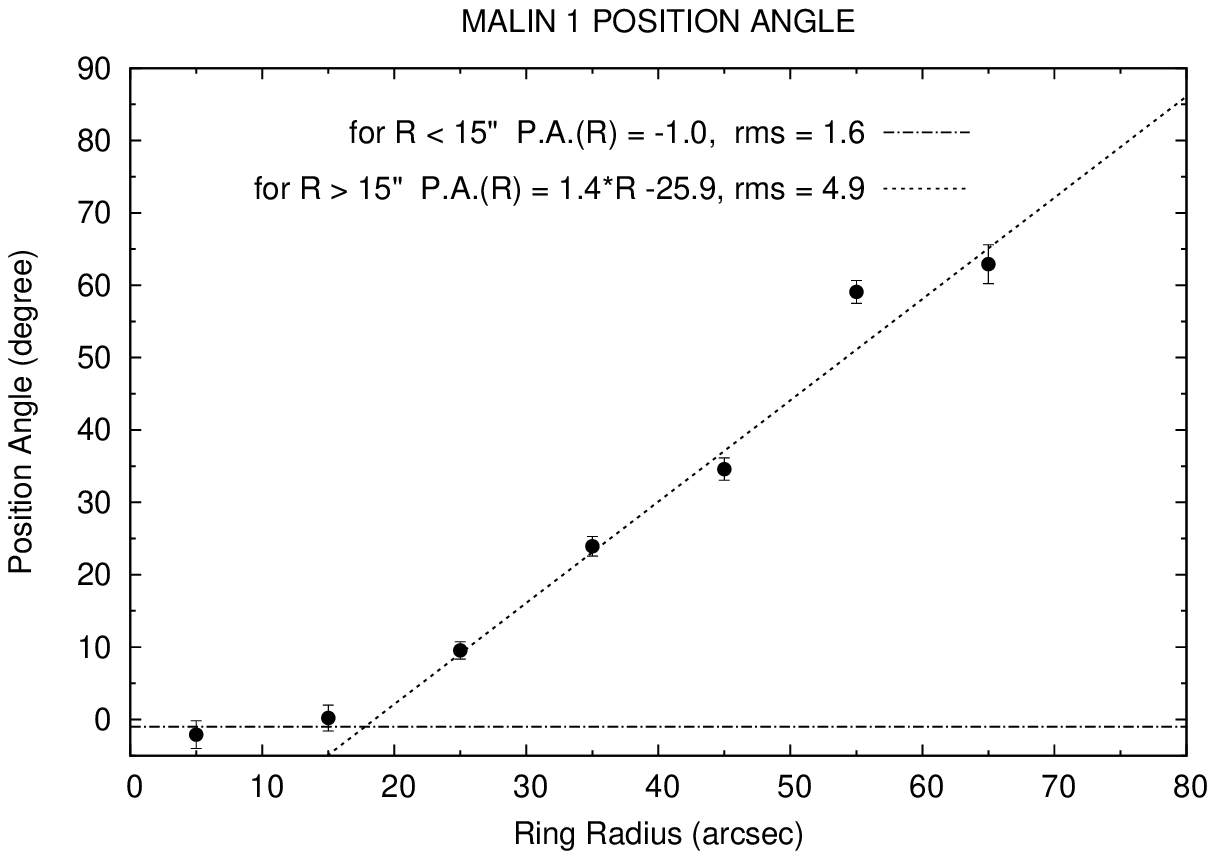}\\
\includegraphics[width=6.5 cm]{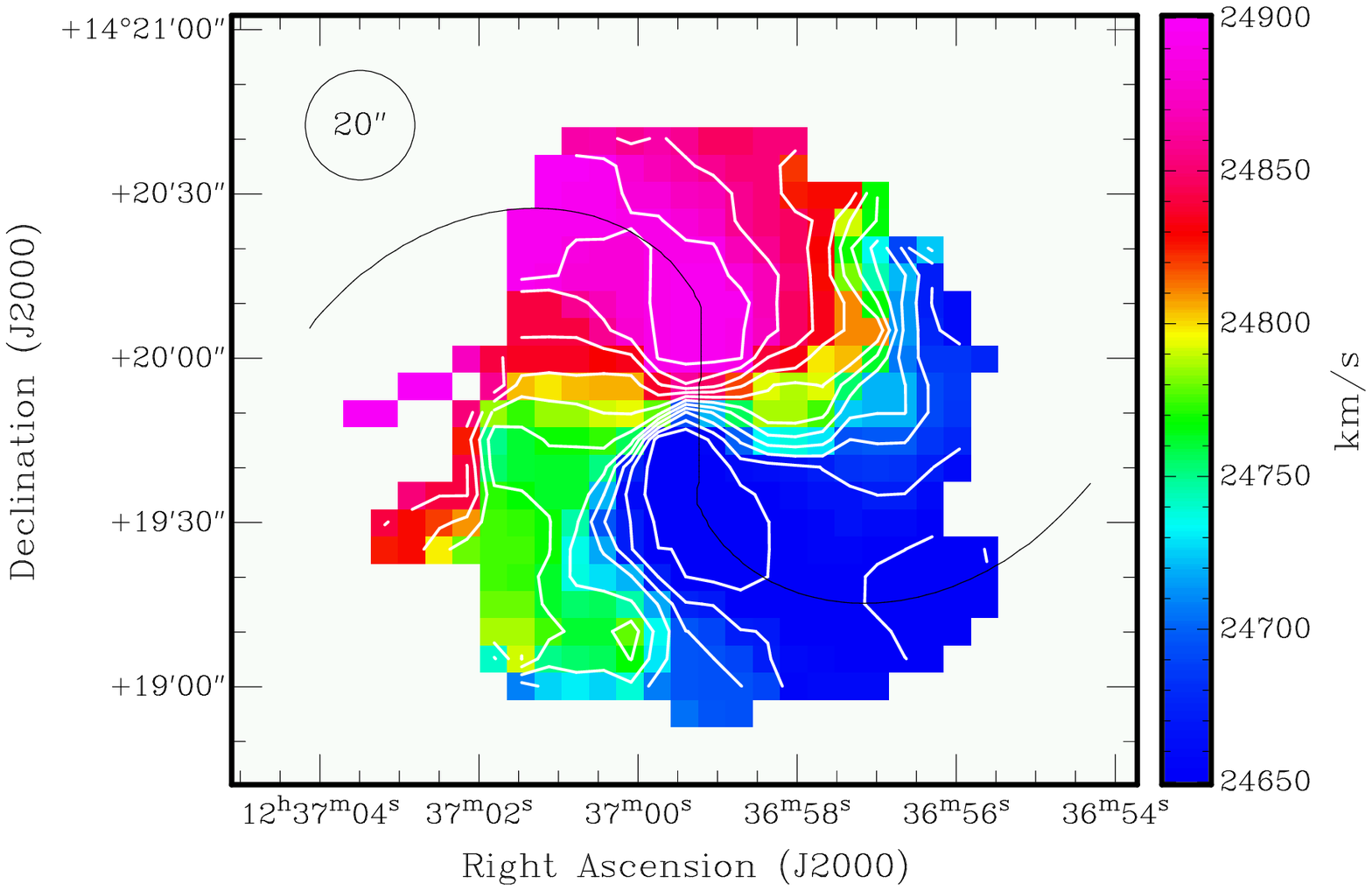}
\end{array}$
\end{minipage}
\caption{\textit{Left Panel}: Position-Velocity diagram for Malin 1 along the Line Of Nodes (LON). The white squares show the projected rotation curve. Full lines show iso-emission contours from 1$\sigma$ to 8$\sigma$ with steps of 1$\sigma$, where 1$\sigma = 0.24$ mJy/beam. Dashed lines show contours at -1$\sigma$ and -2$\sigma$. \textit{Top-Right Panel}: Position angle of the major axis for Malin~1 as a function of radius. The dotted and dash-dotted lines show the fits to the tilted-ring results. \textit{Bottom-Right Panel}: Malin 1 Velocity Field. Iso-velocity contours range from $24647$ km~s$^{-1}$ to $24887$ km~s$^{-1}$ with steps of $24$ km~s$^{-1}$ (2 times the channel spacing). The black line shows the LON.}
\label{fig:pvMalin1}
\end{figure*}

\subsection{\label{RC} Rotation curves}

Kinematical parameters and rotation curves were derived by fitting a tilted-ring model to the observed velocity fields \citep{Begeman1987}. We assumed purely circular motions and described the galaxy by a set of concentric rings. We used all the points of the velocity fields to maximize the statistics of the least-square fit. The points were weighted by $cos^{2}(\theta)$, where $\theta$ is the azimuthal angle in the plane of the galaxy.

As a first step, we determined the kinematical centre $(\alpha_{\rm{kin}}, \delta_{\rm{kin}})$ and the mean systemic velocity $V_{\rm{sys}}$. Then, we kept $(\alpha_{\rm{kin}}, \delta_{\rm{kin}})$ and $V_{\rm{sys}}$ fixed and determined the position angle $P.A.$ as a function of  radius. Subsequently, we tried to determine the inclination angle $i$. Unfortunately, because of the low spatial resolutions, inclination angles and rotational velocities are strongly correlated and no independent determination of $i$ was possible. Therefore, we kept the inclination angles constant, adopting values derived from optical observations. We used $i = 38^{\circ}$ for Malin~1 \citep{MooreParker2006} and $i = 58^{\circ}$ for NGC~7589 \citep{Pickering1997}. Inclinations derived from the total \hi maps are consistent with those values. Finally, we determined the circular velocity $V_{\rm{c}}$ in each ring, keeping all the other parameters fixed. Our results are listed in table \ref{tab:kinparams} and \ref{table:RC}.

In the tilted-ring analysis of NGC~7589, we used a ring width of $20'' \sim 1$ beam. For the kinematical centre, we obtained values which are consistent within the errors with the optical ones. The values obtained for $V_{\rm{sys}}$ and $P.A.$ approximately agree with those of \citet{Pickering1997}.

For Malin~1, we used a ring width of $10'' \sim 1/2$ beam. To determine the kinematical centre, we used the points inside $40'' \sim 62$ kpc. The derived values agree within the errors with the optical ones. The value of $V_{\rm{sys}}$ agrees within the errors with that derived by \citet{Pickering1997}. The $P.A.$ for Malin~1 is not constant. For the inner $15'' \sim 24$ kpc we found $-1^{\circ}.0 \pm 1^{\circ}.6$ and for the outer radii a linear increase up to $65^{\circ}.1 \pm 4^{\circ}.9$ at $R = 65''$ (see the right panels of figure \ref{fig:pvMalin1}). For comparison, \citet{Pickering1997} let the $P.A.$ vary from $0^{\circ} \pm 10^{\circ}$ to $34^{\circ} \pm 3^{\circ}$. To derive the rotation curve of Malin 1, we used a ring width of $16'' \sim 2/3$ beam. In this way, the points of the rotation curve are almost independent.

\begin{figure*}
\centering
\includegraphics[width=18 cm]{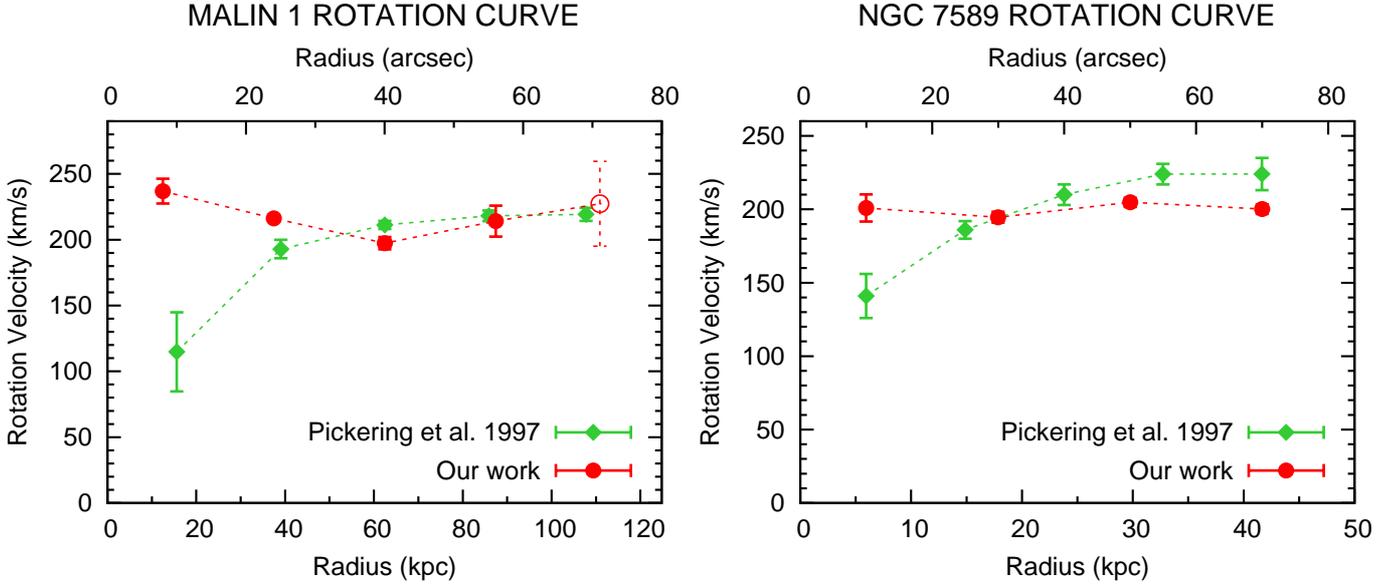}
\caption{\hi rotation curves for Malin~1 (\textit{left}) and NGC~7589 (\textit{right}). Diamonds show the rotation curves derived by \citet{Pickering1997}, filled circles show our new rotation curves. The rotation curve of Malin~1 from \citet{Pickering1997} has been rescaled to $i = 38^\circ$, adopted in our analysis. The open circle was derived only from the approaching half of Malin~1.}
\label{fig:vrot}
\end{figure*}

The errors $\sigma_{V_{\rm{c}}}$ on the circular velocities have been estimated as $\sigma_{V_{\rm{c}}}^{2} = \sigma_{\rm{fit}}^{2} + \sigma_{\rm{asym}}^{2}$, where $\sigma_{\rm{fit}}$ is the formal error given by the fit and $\sigma_{\rm{asym}}$ is an additional uncertainty due to asymmetries between approaching and receding halves. Following \citet{Swaters1999}, we define $\sigma_{\rm{asym}} = (V_{\rm{c, app}} - V_{\rm{c, rec}})/4$, where $V_{\rm{c, app}}$ and $V_{\rm{c, rec}}$ are respectively the approaching and receding rotation curves. Thus, we are assuming that the differences between the rotation curve from the entire velocity field and those for the approaching or receding side correspond to a deviation of $2 \sigma_{\rm{asym}}$. Other uncertainties come from possible variations of the inclination angle. For example, for a change of $i$ equal to 3$^{\circ}$, the rotation curve varies by about 14 km s$^{-1}$ for Malin~1 and 5 km s$^{-1}$ for NGC~7589.

Figure \ref{fig:vrotAppRec} compares the rotation curves derived separately for the approaching and receding sides. For NGC~7589, there is a high degree of symmetry between the two halves. For Malin 1, the symmetry is good out to $R \leq 48'' \sim 75$ kpc. At $R = 56'' \sim 87$ kpc there is a difference of $\sim 45$ km s$^{-1}$ between the two halves, whereas the last point at $R = 72'' \sim 112$ kpc has been derived only for the approaching half. It has a large uncertainty and will be neglected.

Figure \ref{fig:pvMalin1} shows the line of nodes of Malin~1 overlaid onto the velocity field (bottom right panel) and the corresponding position-velocity diagram extracted from the datacube (left panel). The position-velocity diagram is not completely symmetric: the approaching side is more extended than the receding one (see also figure \ref{fig:vrotAppRec}). The projected rotation curve (white squares) closely follows the gas distribution. In particular, in contrast with figure 7 of \citet{Pickering1997}, our first point of the rotation curve corresponds to the peak of the emission.

\subsection{Comparison between the new and the old rotation curves}

\begin{figure*}
\centering
\includegraphics[width=17cm]{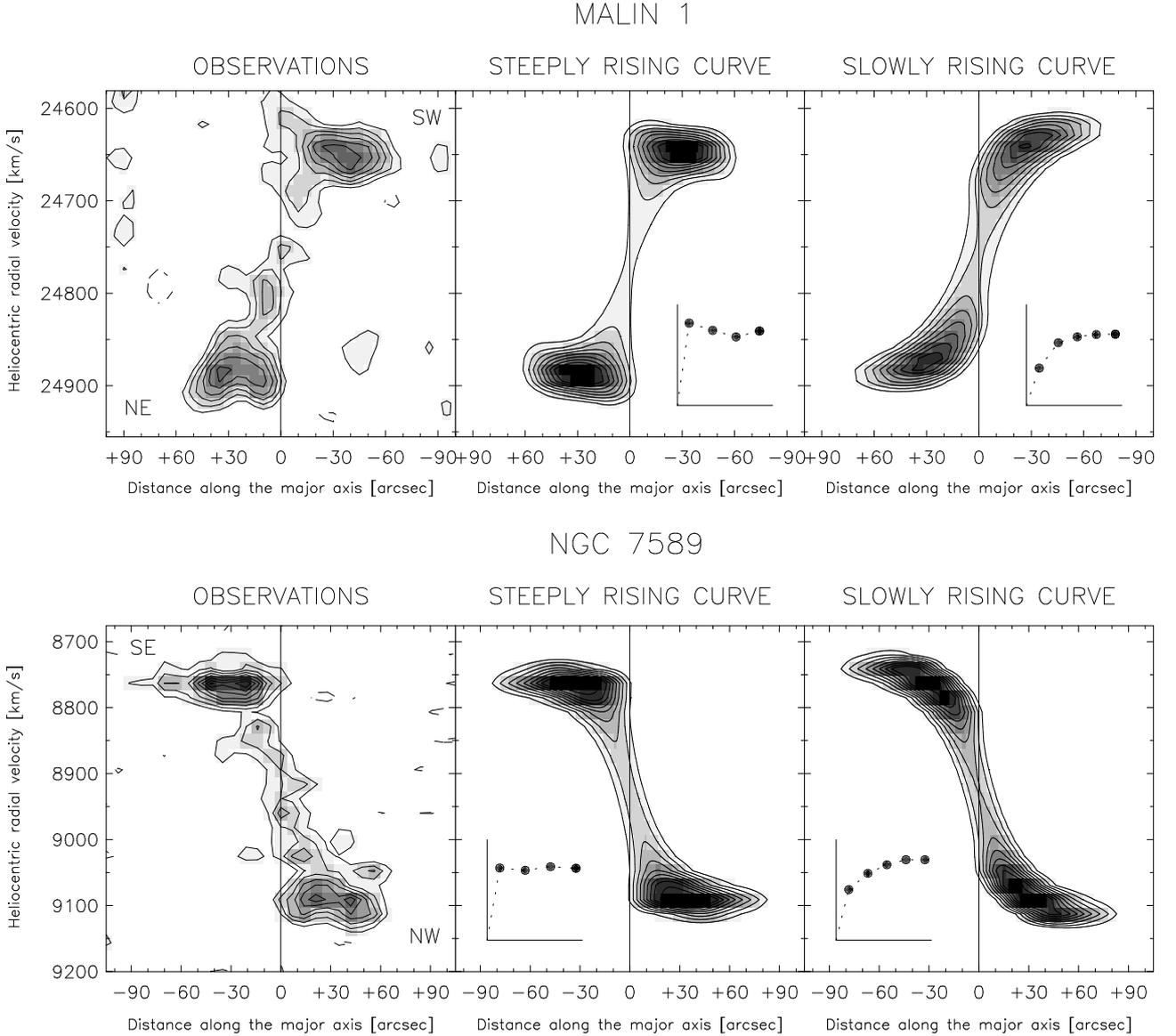}
\caption{Position-Velocity diagrams extracted from the observations and the model datacubes. \textit{From left to right}: observations; model using the steeply rising rotation curve and the other parameters derived in this work; model using the slowly rising rotation curve and the parameters derived by \citet{Pickering1997}. For Malin~1 (\textit{top}) we took a slice along $P.A. = 35^{\circ}$, for NGC~7589 (\textit{bottom}) along $P.A. = 305^{\circ}$. Full lines show iso-emission contours from 2$\sigma$ to 9$\sigma$ with steps of 1$\sigma$. Dashed lines show contours at -2$\sigma$. For Malin 1 (\textit{top}) $\sigma = 0.24$ mJy/beam, for NGC 7589 (\textit{bottom}) $\sigma = 0.35$ mJy/beam.}
\label{fig:pvModels}
\end{figure*}

Figure \ref{fig:vrot} compares our new rotation curves with those derived by \citet{Pickering1997}.
The main difference is that our rotation curves rise more steeply and remain approximately flat out
to the last measured points. As pointed out in section \ref{velo}, the slow rises of the old rotation curves are due to the strong beam smearing effects in the inner regions. In the next section, we construct 3-dimensional galaxy models based on the rotation curves
from  \citet{Pickering1997} and those derived here and demonstrate that the latter are the correct ones.

\begin{table}[thbp]
\caption{Rotation Curves for Malin~1 and NGC~7589.}
\begin{center}
\begin{tabular}{c c c c | c c c c}
\hline
\hline
\multicolumn{4}{c}{MALIN 1} & \multicolumn{4}{c}{NGC 7589} \\
\multicolumn{2}{c}{Radius} & $V_{\rm{c}}$  & $\sigma_{V_{\rm{c}}}$ & \multicolumn{2}{c}{Radius} & $V_{\rm{c}}$  & $\sigma_{V_{\rm{c}}}$\\
($''$) & (kpc)             & (km/s)        & (km/s) & ($''$) & (kpc)             & (km/s)        & (km/s)\\
\hline
8.0 & 12.5  & 236.9 & 9.4 & 10.0 & 5.9  & 200.9 & 9.3 \\
24.0 & 37.5  & 216.2 & 2.6 & 30.0 & 17.7 & 194.6 & 3.7 \\
40.0 & 62.5  & 197.4 & 4.7 & 50.0 & 29.5 & 204.8 & 3.2 \\
56.0 & 87.5  & 214.1 & 11.8 & 70.0 & 41.3 & 200.1 & 3.3 \\
\hline
\hline
\end{tabular}
\tablefoot{
Columns (1) and (2): ring radius in arcsec and kpc. Column (3): circular velocity. Column (4): error on the circular velocity as defined in the text.
}
\label{table:RC}
\end{center}
\end{table}

In the rotation curve of NGC 7589, our last two points lie below those of \citet{Pickering1997}.
The kinematical models described in section \ref{sub:models} show that \citet{Pickering1997}
over-estimated the rotation curve at large radii, but the reason is not clear.

\subsection{\label{sub:models}Kinematical models}

In order to verify the correctness of the new rotation curves, 3-dimensional galaxy models were built. We used the GIPSY task GALMOD, that creates model datacubes by describing a galaxy disk with a set of rings. Each ring is characterized by: the kinematical parameters $(x_{0}, y_{0})$, $V_{\rm{sys}}$, $P.A.$ and $i$, the rotation velocity $V_{\rm{c}}$, the gas velocity dispersion $\sigma_{\rm{\hi }}$, the face-on gas column density $\Sigma_{\rm{\hi }}$ and the scale-height $h_{z}$.

\begin{figure*}
\centering
\includegraphics[width=17cm]{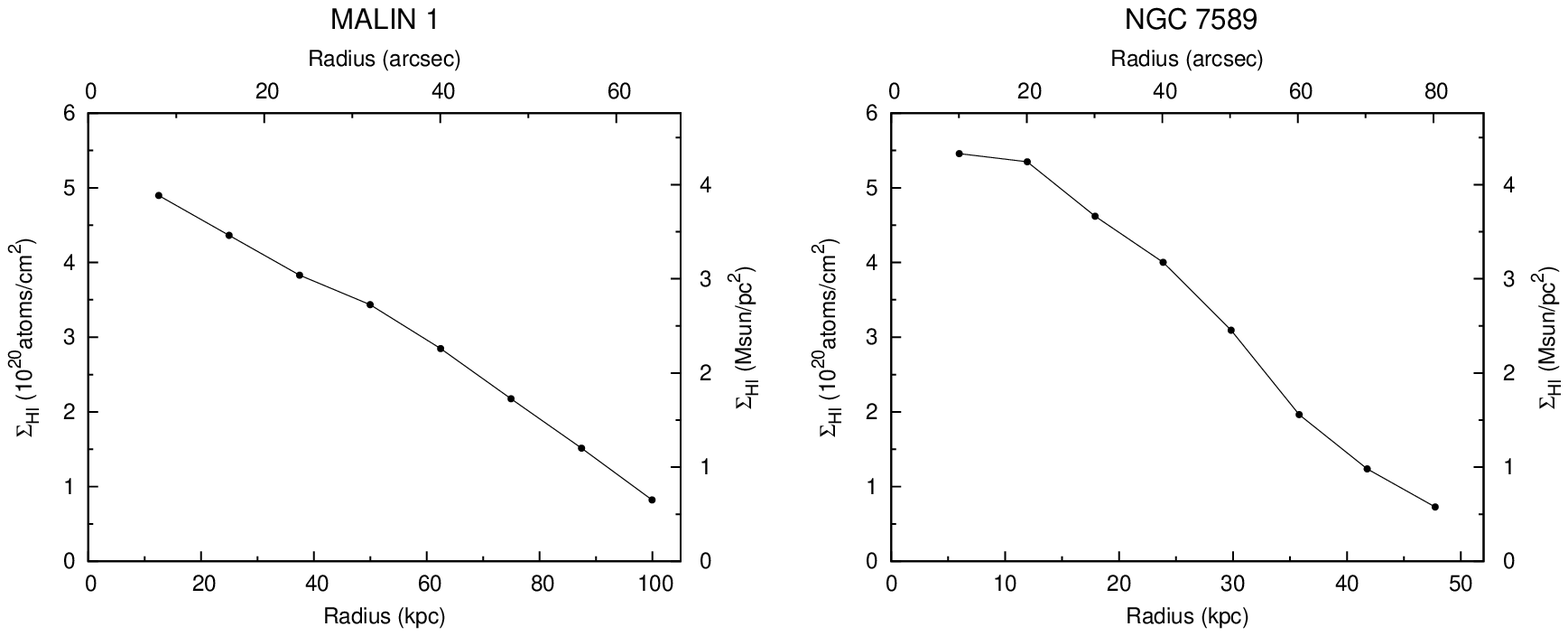}
\caption{HI surface density profile for Malin~1 (\textit{left}) and NGC~7589 (\textit{right}), corrected for inclination and cosmological dimming.}
\label{fig:HIprofile}
\end{figure*}

We created two different sets of models. The first set was built out of our new kinematical parameters and rotation curves. These are the ``steeply rising curve'' models. The second set was built out of the kinematical parameters and rotation curves derived by \citet{Pickering1997}. These are the ``slowly rising curve'' models.
The \hi  surface density profiles were derived from the total \hi  maps, azimuthally averaging over a set of ellipses, determined by the assumed kinematical parameters. For the \hi  vertical distribution, we assumed an exponential profile $\exp(-z/h_{z})$, with $h_{z} = 200$ pc. For the gas velocity dispersion, we used $\sigma_{\rm{\hi }} = 8$ km s$^{-1}$ over the entire disk. The models are almost indistinguishable for differences in $\sigma_{\rm{\hi }}$ of $\pm 4$ km s$^{-1}$. All the models were smoothed to the spatial resolution of the observations, thus they also reproduce beam-smearing effects.

In figure \ref{fig:pvModels}, we compare position-velocity diagrams from the models and the data. For NGC~7589, we took a slice through the datacubes along a position angle of $305^{\circ}$. This is a mean value between our result of $307^{\circ}.7$ and that of $302^{\circ}$ by \citet{Pickering1997}. For Malin~1, we took a slice along a position angle of $35^{\circ}$. This is a mean value for the major axis.

For both galaxies, the observed datacubes (left panels of figure \ref{fig:pvModels}) show the presence of \hi emission at high rotational velocities near the galaxy centre ($R \lesssim 20''$). The ``slowly rising curve'' models (right panels) do not reproduce such emission, as the \hi is spread from low rotational velocities near the centre to high rotational velocities at large radii. On the contrary, in the ``steeply rising curve'' models (middle panels) the \hi emission is concentrated at high rotational velocities, as seen in the data. Both models show tails of emission toward the systemic velocity due to beam-smearing effects. Overall, the observed data are reproduced better by the ``steeply rising curve'' models based on our new results than by the ``slowly rising curve'' models based on those by \citet{Pickering1997}.

We also note that the ``slowly rising curve'' model of NGC~7589 at large radii (R $\gtrsim 50''$) exhibits \hi emission at rotational velocities higher than those observed. This demonstrates that the rotation curve by \citet{Pickering1997} is over-estimated in the outer regions.

\section{\label{tre}Mass Models}

The steeply rising rotation curves found for Malin~1 and NGC~7589 suggest that GLSB galaxies have a dynamical behaviour more similar to a HSB than to a LSB galaxy. To determine the relative contributions of luminous (gas and stars) and dark matter to the gravitational potential, we built mass models following \citet{Begeman1987}.

\subsection{Gas and stars}

The contribution of the gaseous disk was computed using the surface density profiles derived from the total \hi maps (figure~\ref{fig:HIprofile}). 
These were multiplied by a factor $(1+z)^{4}$ to correct for cosmological dimming and by a factor 1.4 to take into account the presence of Helium. 
Molecular and ionized gas were not explicitly considered in the mass model. 
However, since they are usually distributed as the stellar component, their contribution is reflected in a small increase of $M_{*}/L$.
Consistently with the models built in chapter \ref{sub:models}, we assumed an exponential vertical distribution with a scale height of 200 pc.

To compute the contribution of the stellar component, we used surface brightness profiles from the literature. 
For the vertical distribution of the stellar disk, we assumed $Z(z) = sech^{2}(z/z_{0})/z_{0}$ \citep{vanderKruit1981a, vanderKruit1981b}, with $z_{0} = 300$~pc.

\begin{figure*}
\centering
\includegraphics[width=17cm]{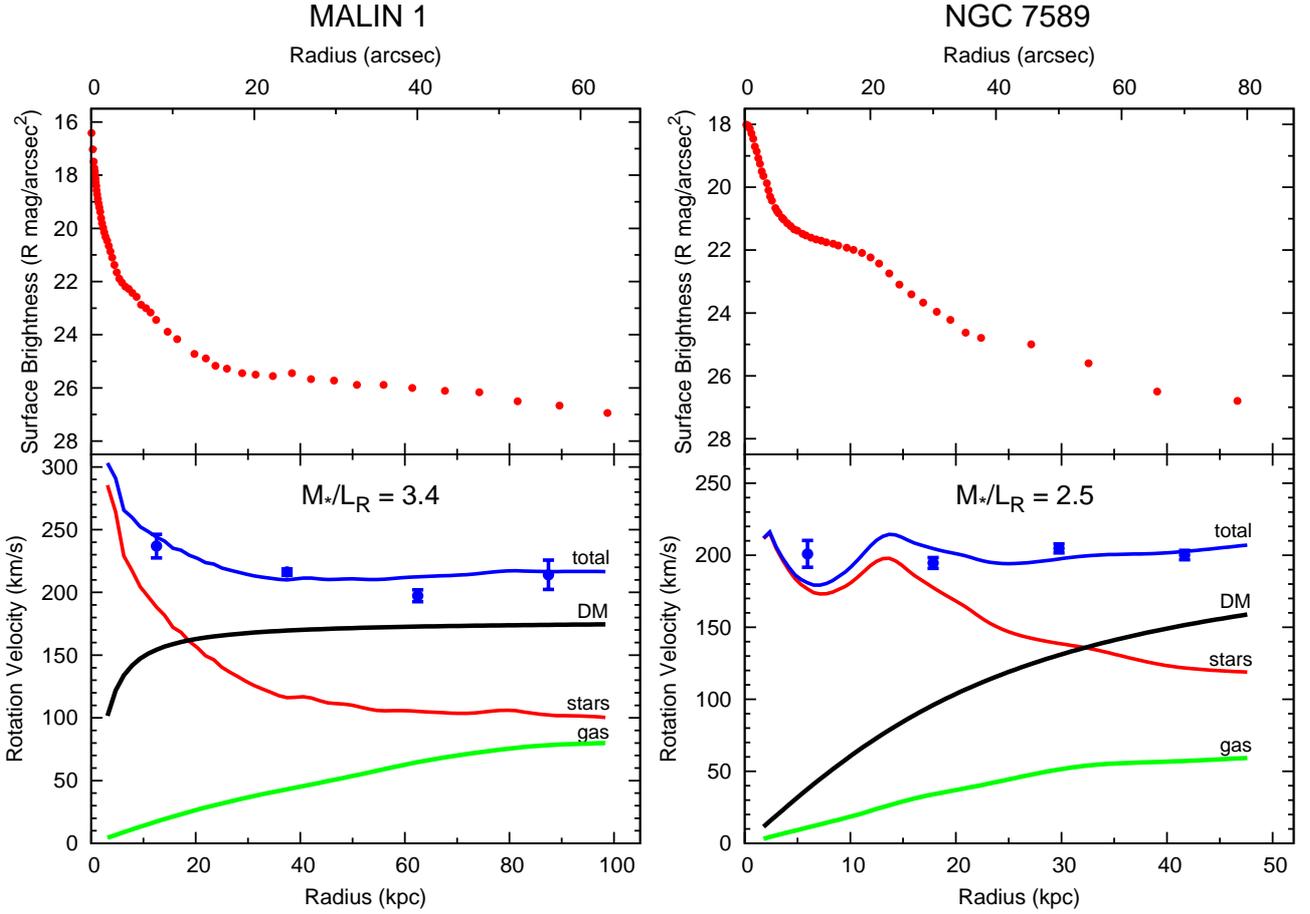}
\caption{\textit{Upper Panels}: R-band surface brightness profiles for Malin~1 (\textit{left}) and NGC~7589 (\textit{right}). See text for details. \textit{Lower Panels}: rotation curve 
decompositions for Malin 1 (\textit{left}) and NGC 7589 (\textit{right}). Dots show the observed rotation curves. The curves show the contributions due to gas, stars and dark matter and the resulting total rotation curve.}
\label{fig:deco1}
\end{figure*}

For Malin~1, we used the I-band profile from \citet{Barth2007} for $R \lesssim 10$ kpc and the R-band profile from \citet{MooreParker2006} for $R \gtrsim 10$ kpc. 
We estimated the R - I colour of the inner galaxy regions using the SDSS data and the relations from \citet{Fukugita1996}. We found $r' - i' = 0.46$, which corresponds to $\rm{R - I} = 0.70$. 
Using this colour, the match between the two profiles is very good (see the upper-left panel of figure \ref{fig:deco1}).
As noted by \citet{Barth2007}, Malin~1 is a ``normal'' SB0/a galaxy, embedded in an extended LSB disk. This double structure is clearly shown by our combined luminosity profile. 
The HSB disk extends out to about 20~kpc, where a photometrically distinct LSB disk appears.

For NGC~7589, we used the R-band profile by \citet{Galaz2006} for $R \lesssim 25$ kpc and the R-band profile by \citet{Pickering1997} for $R \gtrsim 25$ kpc.
In the inner regions, where both profiles are available, there is a systematic difference, maybe due to the different calibration of the observations.
Therefore, we applied a correction of 0.2 mag to the points given by \citet{Pickering1997}, by requiring that the two profiles match in the inner regions.
This correction is within the 1$\sigma$ uncertainties in the photometric measurements.
The result is shown in the upper-right panel of figure \ref{fig:deco1}.
Inside $\sim 20$ kpc, the luminosity profile is typical of a HSB galaxy with a bulge ($R \lesssim 5$ kpc), a lens component ($R \sim 5-13$ kpc) and a HSB exponential disk ($R \sim 13-20$ kpc).
In the outer regions, a photometrically distinct LSB disk extends out to $\sim 50$ kpc.

\begin{figure*}
\centering
\includegraphics[width=17cm]{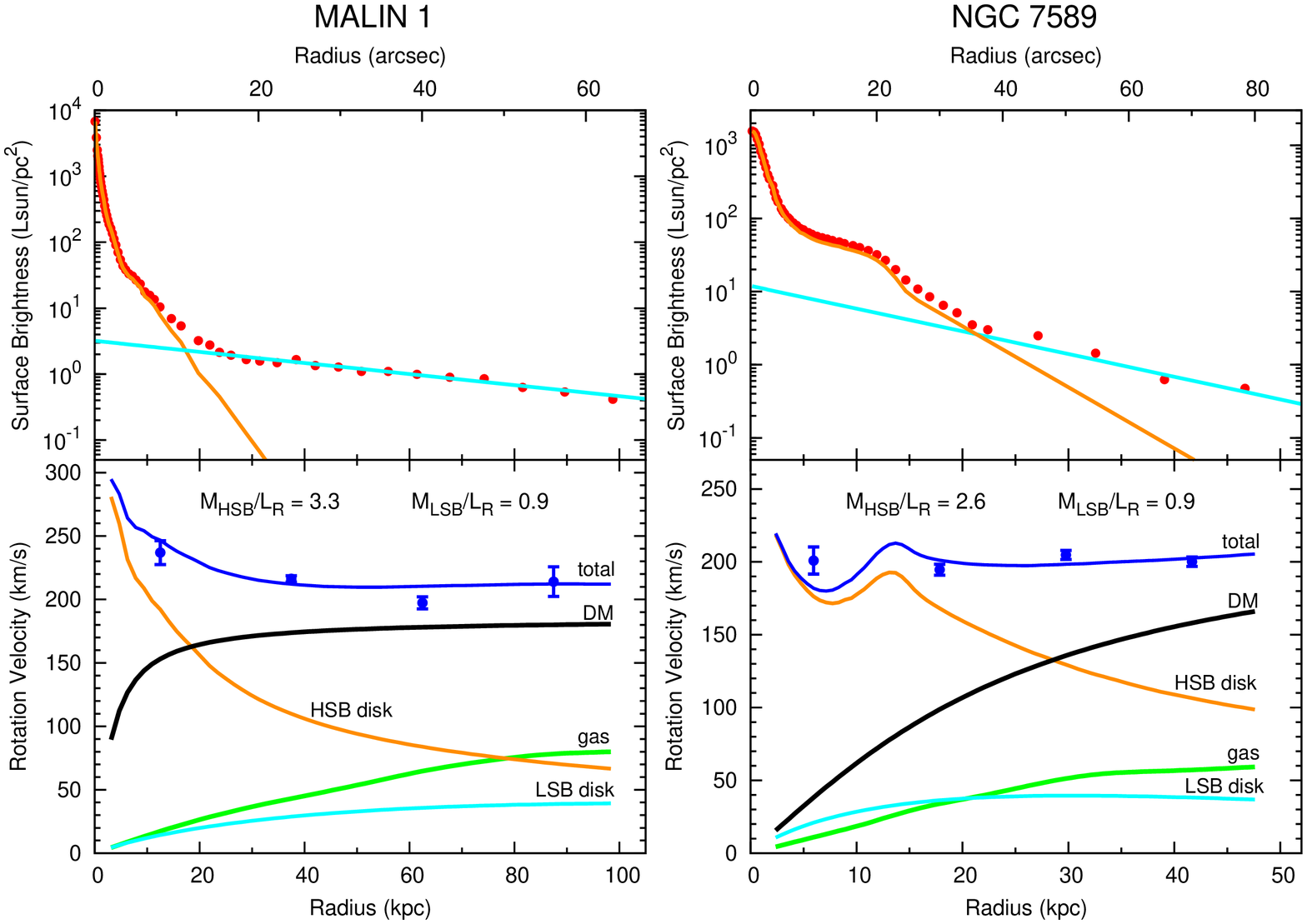}
\caption{\textit{Upper Panels}: R-band surface brightness profiles for Malin~1 (\textit{left}) and NGC~7589 (\textit{right}). The curves show the HSB-LSB decompositions. See text for details. \textit{Lower Panels}: rotation curve decompositions for Malin~1 (\textit{left}) and NGC~7589 (\textit{right}). Dots show the observed rotation curves. The curves show the contributions from the HSB and the LSB stellar disks, the gaseous disk, the dark matter halo and the resulting total rotation curve.}
\label{fig:decoHL}
\end{figure*}

\subsection{Dark matter}

For the dark matter distribution, we assumed pseudo-isothermal haloes described by equation:
\begin{equation} \label{eq:ISO}
 \rho_{\rm{ISO}}(r) = \frac{\rho_{\rm{0}}}{1 + (r/r_{\rm{C}})^{2}},
\end{equation}
where $\rho_{\rm{0}}$ is the central density and $r_{\rm{C}}$ is the core radius. $\rho_{\rm{0}}$ and $r_{\rm{C}}$ are free parameters of the mass models.

We also tried rotation curve decompositions with NFW dark matter haloes \citep{Navarro1995} but found that the results are not significantly different.

\subsection{Rotation curve decompositions}

The \hi rotation curves have much lower spatial resolution than the surface brightness profiles (and their resulting dynamical contributions). Thus, a least-square fit done only on the observed points of the rotation curves would neglect some features present in the luminosity profiles. To avoid this problem, we have linearly interpolated the rotation curves and performed a fit with steps of $1''$. Errors on the rotation curves were used as weights in the fit.

Figure \ref{fig:deco1} shows the ``maximum disk'' decompositions of the rotation curves.
The resulting mass-to-light ratios are $M_{*}/L_{\rm{R}} = 3.4$ for Malin~1 and $M_{*}/L_{\rm{R}} = 2.5$ for NGC~7589 (Table \ref{table:darkmatter}).
These values are somewhat uncertain due to the lack of resolution in the central regions.
However, we estimate that the mass-to-light ratio cannot be higher than 5 for Malin\,1 and 3 for NGC\,7589.
Thus, they surely are in the range of values typically found for HSB early-type spirals \citep{Noordermeer2005}.
Under the maximum disk hypothesis, the baryons dominate the galaxy dynamics out to a radius of $R\sim 20$ kpc for Malin~1 and $R\sim 30$ kpc for NGC~7589. In contrast with the results of \citet{Pickering1997}, dark matter does not necessarily dominate everywhere.

\begin{table}[thbp]
\caption{Mass Models for Malin~1 and NGC~7589.}
\begin{center}
\begin{tabular}{l c c}
\hline
\hline
Parameter & MALIN 1 & NGC 7589 \\
\hline
$M_{*}/L_{\rm{R}}$ ($M_{\odot}/L_{\odot}$) & 3.4 & 2.5\\
$\rho_{0}$ ($10^{-3} \; M_{\odot}$ pc$^{-3}$) & $121.4 \pm 93.8$ & $2.3 \pm 0.3$\\
$r_{\rm{C}}$ (kpc) & $2.2 \pm 0.9$ & $19.9 \pm 2.0$\\
$\chi^{2}_{r}$ & 4.6 & 4.0\\
\hline
\hline
$M_{\rm{HSB}}/L_{\rm{R}}$ ($M_{\odot}/L_{\odot}$) & 3.3 & 2.6\\
$M_{\rm{LSB}}/L_{\rm{R}}$ ($M_{\odot}/L_{\odot}$) & 0.9 & 0.9\\
$\rho_{0}$ ($10^{-3} \; M_{\odot}$ pc$^{-3}$) & $73.9 \pm 36.5$ & $2.4 \pm 0.2$\\
$r_{\rm{C}}$ (kpc) & $2.9 \pm 0.8$ & $20.7 \pm 1.6$\\
$\chi^{2}_{r}$ & 3.1 & 2.7\\
\hline
\hline
\end{tabular}
\end{center}
\tablefoot{\textit{Top}: mass models using a single stellar component and maximizing the stellar mass-to-light ratio. \textit{Bottom}: mass models using a HSB-LSB decomposition. The stellar mass-to-light ratio of the LSB disk was fixed at 0.9, whereas that of the HSB disk was maximized. See text for details.}
\label{table:darkmatter}
\end{table}

Both Malin~1 and NGC~7589 show colour gradients, which may indicate different stellar populations. 
\citet{Bothun1987} studied the luminous ``blobs'' in the outer disk of Malin~1 (see figure \ref{fig:Maps}) and found a mean $\rm{B - V}$ colour of 0.37, whereas in the inner parts (R $\lesssim 20''$) the mean $\rm{B - V}$ colour is 0.90 \citep{Impey1989}. 
\citet{Galaz2006} traced the ${\rm B-R}$ colour profile of NGC~7589 out to $R \sim 20$ kpc and found complex trends, but in general the galaxy tends to become bluer at large radii.
It makes sense, therefore, to decompose the luminosity profiles and use two different $M_{*}/L_{\rm{R}}$ for the inner and the outer regions.

We separated the LSB disk by fitting an exponential profile to the points beyond $R \gtrsim 20$ kpc. The resulting disk parameters are: $\mu_{R}(0) = 24.7$ mag arcsec$^{-2}$, $h = 51.7$ kpc for Malin~1 and $\mu_{R}(0) = 23.3$ mag arcsec$^{-2}$, $h = 14$ kpc for NGC~7589. The HSB component was isolated by subtracting the LSB disk from the total photometric profile (see top panels of Fig. \ref{fig:decoHL}).
We maximized the contribution of the HSB component and found $M_{\rm{HSB}}/L_{\rm{R}} = 3.3$ for Malin~1 and $M_{\rm{HSB}}/L_{\rm{R}} = 2.6$ for NGC~7589, very similar to the previous values.
Maximizing also the contribution of the LSB disk, the outer parts of the rotation curve can be reproduced  without the need for a dark matter halo. However, the resulting $M_{\rm{LSB}}/L_{\rm{R}}$ ratios would be unrealistically high ($\sim$25 for Malin~1 and $\sim$15 for NGC~7589), as it is usually found by maximizing the disks of ``ordinary'' LSB galaxies (e.g. \citealt{deBlok2001}). 
Therefore, we estimated $M_{\rm{LSB}}/L_{\rm{R}} = 0.9$ using the relations of \citet{Bell2003} and assuming $\rm{B - V} = 0.44$. 
The latter value is the mean colour found by \citet{McGaugh1994a} for a sample of bulgeless LSB disks. 
The results of the fits are listed in table \ref{table:darkmatter} and shown in the bottom panels of figure~\ref{fig:decoHL}.

We also performed a standard bulge-disk decomposition for the inner HSB part and found a degeneracy of the mass-to-light ratios of bulge and disk.
Different combinations of $M_{\rm{bulge}}/L_{\rm{R}}$ and $M_{\rm{disk}}/L_{\rm{R}}$ provide comparable fits, as the rotation curves are not resolved in the inner regions.
By maximizing both bulge and disk we found that their contributions to the first point of the rotation curve are approximately equal. Therefore, in this ``maximum light'' hypothesis, the inner steep rise is not explained only with the bulge, but also the disk is dynamically important.
This further suggests that Malin~1 and NGC~7589 have an inner ``normal'' HSB disk (see also \citealt{Barth2007}).

Following \citet{McGaugh2005}, we used the derived values of $M_{*}/L_{\rm{R}}$ and $M_{\rm{\hi}}$ to check the position of Malin~1 and NGC~7589 on the baryonic Tully-Fisher relation. We found that both galaxies follow the relation within the observed scatter.

\subsection{\label{MOND}MOND}

The modified newtonian dynamics (MOND) was proposed by \citet{Milgrom1983} as an alternative to dark matter. This theory modifies the force law at accelerations $a$ lower than a critical value $a_{0}$ and naturally explains the flatness of rotation curves (see the review by \citealt{Sanders2002}).

GLSB galaxies are ideal systems to test MOND. Firstly, they are very diffuse and extended, thus the test is possible down to very low values of $a/a_{\rm{0}}$ (deep MONDian regime). Secondly, they have massive gaseous disks. In the MONDian framework, this gives an important contribution to the galaxy dynamics, which is not subject to uncertainties as that of stars (i.e. the value of $M_{*}/L$). For the MOND fit to the rotation curves, we followed \citet{Begeman1991} and assumed $a_{0} = 3000$ km$^{2}$ s$^{-2}$ kpc$^{-1}$ \citep{Bottema2002}. 
Two different M/L ratios for the inner and outer regions of the galaxies have been used as free parameters in the fit.

\begin{figure}
\centering
\includegraphics[width=8.5 cm]{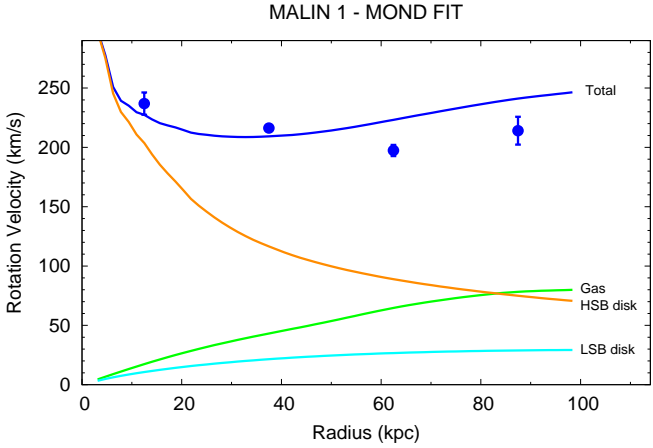}
\caption{MOND applied to Malin~1 using two different $M_{*}/L_{\rm{R}}$ for the inner HSB region ($M_{\rm{HSB}}/L_{\rm{R}} = 3.7$) and the outer LSB disk ($M_{\rm{LSB}}/L_{\rm{R}} = 0.5$). Dots show the observed rotation curve. Lines show the Newtonian contributions due to the gaseous, the HSB and the LSB disk and the result of the MOND fit. See text for details.}
\label{fig:MOND2c}
\end{figure}

For NGC~7589, we obtained a good fit by using a bulge-disk decomposition. The stellar mass-to-light ratios predicted by MOND are acceptable: $M_{\rm{bulge}}/L = 4.7$ and $M_{\rm{disk}}/L = 1.3$.
For Malin~1, it is preferable to use the HSB-LSB decomposition shown in figure \ref{fig:decoHL}, but it is not possible to leave both $M_{\rm{HSB}}/L$ and $M_{\rm{LSB}}/L$ free.
Indeed, the fit would require a negative value for $M_{\rm{LSB}}/L_{\rm{R}}$.
This suggests that the presence of the extended LSB disk, combined with the large amount of gas, creates problems for MOND.
We fixed $M_{\rm{LSB}}/L_{\rm{R}}$ and used only $M_{\rm{HSB}}/L_{\rm{R}}$ as a free parameter. Figure \ref{fig:MOND2c} shows the result for $M_{\rm{LSB}}/L_{\rm{R}} = 0.5$.
Using the relations by \citet{Bell2003}, this corresponds to an extremely blue colour for the outer disk, i.e. B-V = 0.2. Higher values of $M_{LSB}/L_{R}$ worsen the fit, whereas lower values only slightly improve it, as the gas becomes dominating in the outer regions.
A bulge-disk decomposition, as that used for NGC~7589, does not improve the fit. Higher values of the critical acceleration, as $a_{0} = 3700$ km$^{2}$ s$^{-2}$ kpc$^{-1}$ \citep{Begeman1991} or $a_{0} = 4000$ km$^{2}$ s$^{-2}$ kpc$^{-1}$ \citep{McGaugh2004}, worsen the fit. Interpolation functions different from the standard one, as those proposed by \citet{Famaey2005} or \citet{Milgrom2008}, give slightly worse fits.

Although the rotation curve of Malin 1 is not of high quality, the discrepancies between the observed velocities and the MOND prediction are significant as they are equal to $\sim 25$ km s$^{-1}$, i.e. $\sim 12-13 \%$.
However, the warp in the outer disk of Malin~1 may greatly affect these results. The rotation curve used here was derived by varying the $P.A.$ with radius (see figure \ref{fig:pvMalin1}) but assuming a constant inclination angle of $38^{\circ}$. It was not possible to obtain a better determination of $i$ from the velocity field. A change of the inclination angle in the outer parts of only $6^{\circ}$ from $i=38^{\circ}$ to $i=32^{\circ}$, which cannot be ruled out, would result in a rotation curve in total agreement with the MOND prediction. 
Note that here we have used the simplest formulation of MOND \citep{Milgrom1983}, that may not strictly apply to warped non-axisymmetric disks.

\section{\label{quattro}Discussion}

\subsection{\label{sub:quattrouno}GLSB galaxies and their double HSB-LSB nature}

It is known that there is a dichotomy between HSB and LSB systems (e.g. \citealt{Verheijen1999}). HSB galaxies usually have steeply rising rotation curves and can be described by a maximum disk,
whereas LSB galaxies have slowly rising rotation curves and are thought to be dominated by dark-matter everywhere.
The dynamics of Malin~1 and NGC~7589 is typical of HSB systems. The concentration of mass, indicated by the steeply rising rotation curve, follows the inner concentration of light. For both galaxies, maximum-disk solutions have been found with reasonable values of $M_{*}/L$. Thus, we conclude that in the inner regions either baryons are dominating or, in case dark matter dominates, it must follow the distribution of light. This may be in line with some recent numerical simulations (\citealt{Tissera2009}), which suggest a gravitational effect of baryons on shaping the distribution of dark matter.

Both surface photometry and gas dynamics indicate that Malin~1 and NGC~7589 have a
double structure: an inner ``normal'' spiral galaxy (with a bulge and an HSB disk)
and an outer extended LSB disk.
Is such a double structure present in the other GLSB galaxies as well?
The other two GLSB galaxies studied by \citet{Pickering1997} are F568-6 (Malin~2) and UGC~6614.
F568-6 is kinematically lopsided and difficult to interpret.
UGC~6614 was observed in H$\alpha$ by \citet{McGaugh2001} who found a steeply rising rotation curve
(see their figure~5) and thus a dynamical behaviour similar to that of a HSB system
(see also \citealt{deBlok2001}).
\citet{Walsh1997} studied the GLSB galaxy NGC~289 using optical and radio observations.
They found an inner HSB disk and a steeply rising rotation curve that can be fitted by a
maximum disk with $M_{*}/L_{\rm{I}} = 2.3$. They drew conclusions very similar to ours.
Another example is NGC~5383. This barred spiral galaxy fits in the GLSB class. It has
a LSB disk with $h \sim 9.7$ kpc and a G-band absolute
magnitude of $\rm{M}_{\rm{G}} = 20.6$ \citep{vanderKruit1978}\footnote{The original values
have been rescaled to a distance of 33.7 Mpc, as we assume $H_{0} = 70$ km s$^{-1}$ Mpc$^{-1}$.}.
Its surface brightness profile is very similar to that of NGC~7589
and extends out to $\sim 54$ kpc (see \citealt{Barton1997}).
The HI mass is $M_{\hi} = 5.9 \times 10^{9}$ M$_{\odot}$
and the gas dynamics is typical of a HSB system \citep{Sancisi1979}.

\citet{Sprayberry1995} catalogued another 13 objects as GLSB~galaxies. As argued by \citet{Barth2007},
their inner HSB disk could have been missed due to the low spatial resolution of the optical
observations. However, for some of them, \hi rotation curves were derived. These galaxies are
NGC~5533 \citep{Noordermeer2007}, NGC~5905 \citep{vanMoorsel1982}, NGC~4017 \citep{vanMoorsel1983},
NGC~2770 \citep{GarciaRuiz2002}, UGC~2936 \citep{Pickering1999} and PGC~45080 \citep{Das2007}.
All of them show a dynamical behaviour typical of HSB~systems, with the possible exception of
PGC~45080 that is kinematically lopsided (see figure~11 of \citealt{Das2007}). Therefore, it seems
unlikely that GLSB galaxies constitute a sub-class of the bulge-dominated LSB galaxies
(e.g. \citealt{Beijersbergen1999}, \citealt{Galaz2006}). It is likely, instead, that they are
``normal'' HSB galaxies which possess an outer extended LSB disk.

GLSB galaxies do show some physical properties closer to HSB than to LSB systems.
Usually, LSB galaxies are quite amorphous and rarely host an AGN \citep{Impey1997, Bothun1997}. In contrast, GLSB galaxies can have a well-defined spiral pattern \citep{Beijersbergen1999} and show nuclear activity with the same probability as HSB spirals \citep{Schombert1998}. Moreover, LSB galaxies are deficient in molecular gas (e.g. \citealt{O'Neil2003a, Matthews2005}), whereas CO emission has been detected in some GLSB galaxies (e.g. \citealt{O'Neil2004a, Das2006}).

At this point it is natural to ask how common it is for HSB galaxies to have diffuse outer
LSB stellar disks.
This question has been addressed in the past (see e.g.\ \citealt{Bosma1993} and references therein)
and examples were found such as NGC~3642 \citep{VerdesMontenegro2002}.
More recently, \citet{Erwin2008} studied a sample of 66 S0-Sb galaxies and found that $\sim 24 \%$ of them show
``anti-truncated'' disks, i.e. the surface brightness profile of the disk
can be described by two
exponentials, the outer one with larger scalelength than the inner one.
Such ``up-bending'' luminosity profiles have also been observed
in late-type spiral galaxies \citep{Hunter2006, Pohlen2006}.
GLSB galaxies may be extreme examples of anti-truncated early-type HSB spirals.
Moreover, GALEX observations have recently revealed extended XUV-disks around
HSB spirals \citep{Thilker2007}. Also some GLSB galaxies have
been observed with GALEX \citep{Boissier2008} and, at such wavelengths,
they resemble the XUV-disk galaxies of \citet{Thilker2007} (see in particular Malin~1, Malin~2 and NGC~7589).

Regarding gaseous disks, it is known that a large number of spiral galaxies, perhaps the majority,
have \hi disks extending far beyond the bright optical ones (e.g. \citealt{Sancisi2008}). The outlying \hi
has typical surface densities of 1-2 $M_{\odot}$ pc$^{-2}$, close to the values measured in GLSB galaxies.
Similar results have been found also for early-type galaxies in the field.
\citet{Oosterloo2007} have observed at 21-cm a sample of 30 gas rich E/S0 galaxies and discovered that
about 2/3 are embedded in rotating \hi disks with masses, densities and sizes comparable to the gaseous
disk of Malin~1. Deep optical observations would be useful to verify whether such spirals and ellipticals
have extended LSB stellar disks associated with the gaseous ones.
If this is the case, these objects could be considered to be in the same class of GLSB galaxies.

\subsection{The formation of GLSB galaxies}

GLSB galaxies are a challenge for theories of galaxy formation.
The main properties to be explained are: i) the double HSB-LSB nature, ii) the
large extent of the outer LSB disks and iii) the regular and symmetric large-scale
kinematics of the latter, as observed in both Malin 1 and NGC 7589 (see figure \ref{fig:vrotAppRec}).
Given the very large orbital periods at the outermost radii (e.g. $t_{\rm{dyn}} \sim 2.8$ Gyr
for Malin~1 and $t_{dyn} \sim 1.4$ Gyr for NGC~7589) it is clear that these outer disks must
have been in place and undisturbed for several Gyrs.

CDM cosmological simulations of galaxy formation tend to produce dense and compact disks, as much angular momentum is lost during the hierarchical assemblage (see \citealt{Kaufmann2007} and references therein).
\citet{Hoffman1992} showed that rare density peaks ($\sim$ 3$\sigma$) in underdense environments (voids) may lead to the formation of a GLSB galaxy. However, GLSB galaxies show the same clustering properties of HSB spirals and are not necessarily associated with voids \citep{Sprayberry1995}.

\citet{Mapelli2008} simulated the evolution of a collisional ring galaxy.
The propagation of the ring leads to the formation of a large and diffuse system,
with structural properties similar to those of GLSB galaxies.
However, these simulations predict slowly
rising rotation curves and therefore a different dynamics from that of GLSB galaxies.

\citet{Noguchi2001} proposed a secular-evolution scenario, starting from a massive barred HSB galaxy. The bar-instability can redistribute the matter and the angular momentum in the disk, but the scale length $h$ cannot increase by more than a factor of 2-2.5 (see also \citealt{Debattista2006}). Therefore, this mechanism cannot explain the formation of an extended LSB disk as that of Malin 1 (with $h \sim 50$ kpc). Moreover, it should explain why the majority of barred early-type spirals do not have an outer LSB disk \citep{Erwin2008}.

Simulations by \citet{Penarrubia2006} investigated the formation of
LSB disks around HSB galaxies by interactions with dwarf companions.
They found that the dwarf galaxies are disrupted by tidal forces and
the debris settle on an extended disk with a nearly exponential profile.
Interestingly, the scale length of the resulting disk depends on the compactness
of the original stellar distribution in the dwarfs and, in extreme cases,
it can be as large as $\sim 50$ kpc. Thus, it would be possible to explain a
wide range of disk structural parameters as those observed in GLSB galaxies.
However, such simulations predict a decrease in the rotation velocity of the outer
stellar disk of about 30-50 km~s$^{-1}$ with respect to the inner one.
Such a decrease is not observed in the \hi rotation curves of GLSB galaxies.

\section{\label{cinque}Conclusions}

We studied the structural and dynamical properties of
two giant low surface brightness galaxies: Malin 1 and NGC 7589.
For both galaxies, we re-analyzed existing 21-cm line observations,
obtained with the VLA by \citet{Pickering1997}, and derived
new rotation curves. The quality of these curves was tested by building
model datacubes. Finally, we decomposed the rotation curves
using available optical surface brightness profiles and
assuming pseudo-isothermal dark matter haloes.
We also studied the predictions of MOND.
Our main results can be summarized as follows:

\begin{enumerate}

\item in contrast with the previous results by \citet{Pickering1997}, both Malin~1 and NGC~7589 have steeply rising rotation curves. From a dynamical point of view, these GLSB galaxies are more similar to early-type HSB than to late-type LSB disks. This agrees with the result of the optical study by \citet{Barth2007}
that described Malin 1 as a normal SB0/a galaxy surrounded by a huge LSB disk.
Both surface photometry and gas dynamics indicate that also NGC 7589 has such a double HSB-LSB structure.

\item In these two GLSB galaxies, the concentration of mass indicated by the steeply rising rotation curve corresponds to the inner concentration of light. The rotation curves of both galaxies can be fitted by a maximum disk with $M_{*}/L_{\rm{R}} \sim 3$. Contrary to the case of the less massive LSB disks where dark matter is believed to dominate everywhere, in the inner regions of GLSB galaxies either the baryons dominate or the dark matter follows the distribution of light.

\item Malin~1 provides a severe test for MOND because of its extraordinary size and high \hi mass. The rotation curve derived here differs significantly from the one predicted by MOND. However, the discrepancies would disappear if the outer disk of Malin~1 were warped and turned to more face-on. Unfortunately, the present observations are not adequate to determine the trend of the inclination angle.
\end{enumerate}

\begin{acknowledgements}
We thank Tim Pickering and his co-authors for kindly
making their HI datacubes available to us. We also thank
Gaspar Galaz and his co-authors for the photometric
data of NGC~7589 and Lesa Moore for the optical image of Malin~1.
We are grateful to Carlo Nipoti and Bob Sanders for
stimulating discussions about MOND and to the referee
Albert Bosma for helpful comments and suggestions.
\end{acknowledgements}

\bibliographystyle{aa}
\bibliography{bibliography6.bib}

\end{document}